\documentclass[aps,prl,reprint,superscriptaddress]{revtex4-2}

\usepackage{color}
\usepackage{soul}
\usepackage{dcolumn}
\usepackage{graphicx}
\usepackage{amsmath}
\usepackage{amssymb}
\usepackage{bm}

\begin{document}
\mbox{}

\title{Inelastic scattering of a photon by a quantum phase-slip}

\author{R. Kuzmin}
\affiliation{Department of Physics, University of Maryland, College Park, Maryland 20742, USA.}
\author{N. Grabon}
\affiliation{Department of Physics, University of Maryland, College Park, Maryland 20742, USA.}

\author{N. Mehta}
\affiliation{Department of Physics, University of Maryland, College Park, Maryland 20742, USA.}

\author{A. Burshtein}
\affiliation{Raymond and Beverly Sackler School of Physics and Astronomy, Tel Aviv University, Tel Aviv 6997801, Israel.}
\author{M. Goldstein}
\affiliation{Raymond and Beverly Sackler School of Physics and Astronomy, Tel Aviv University, Tel Aviv 6997801, Israel.}
\author{M. Houzet}
\affiliation{Univ. Grenoble Alpes, CEA, INAC-Pheliqs, F-38000 Grenoble, France.}
\author{L. I. Glazman}
\affiliation{Department of Physics, Yale University, New Haven, CT 06520, USA.}
\author{V. E. Manucharyan}
\affiliation{Department of Physics, University of Maryland, College Park, Maryland 20742, USA.}

\email[]{manuchar@umd.edu}

\date{\today}
\begin{abstract}
Spontaneous decay of a single photon is a notoriously inefficient process in nature irrespective of the frequency range. We report that a quantum phase-slip fluctuation in high-impedance superconducting waveguides can split a single incident microwave photon into a large number of lower-energy photons with a near unit probability. The underlying inelastic photon-photon interaction has no analogs in non-linear optics. Instead, the measured decay rates are explained without adjustable parameters in the framework of 
a new model of a quantum impurity in a Luttinger liquid. Our result connects 
circuit quantum electrodynamics to critical phenomena in two-dimensional boundary quantum field theories, important in the physics of strongly-correlated systems. The photon lifetime data represents a rare example of verified and useful quantum many-body simulation.

\end{abstract}

\maketitle

\textbf{}

Although photons have zero mass, fundamental laws do not prevent their splitting into more photons as soon as some form of non-linearity is present. Thus, individual  $100~\textrm{MeV}$-photons split in the Coulomb field of heavy nuclei because of vacuum polarization~\cite{akhmadaliev2002experimental} and so do optical photons in non-linear crystals via the process of spontaneous parametric down-conversion~\cite{guerreiro2014nonlinear,Bock16}. However, the splitting probability is extremely low, e.g. it does not exceed $10^{-6}$ per cm of optical crystal, the origin of which can be traced down to the small value of the fine-structure constant. Interactions at the single photon level are known to be dramatically enhanced
in circuit quantum electrodynamics (cQED), owing to both the reduced mode volume of microwave transmission lines and the non-linearity of Josephson junctions~\cite{blais2020quantum}.
Notable achievements include observations of vacuum Rabi~\cite{10.1038/nature02851} and photon number~\cite{schuster2007resolving} splittings, resonance fluorescence~\cite{astafiev2010resonance}, as well as implementations of multi-mode~\cite{sundaresan2015beyond} and ultrastrong coupling regimes~\cite{forn2019ultrastrong, kockum2019ultrastrong}. 
Yet, spontaneous down-conversion remains \textit{improbable} even in cQED. The splitting of photons into two~\cite{bergeal2012two} or three~\cite{chang2020observation} ones was observed only under a strong stimulation of non-linear circuits by a classical field, a process that can be well understood using semi-classical wave mixing equations~\cite{bergeal2010analog}. In stark contrast, we encountered an efficient quantum mechanism of photon-photon interaction in high-impedance superconducting waveguides. Without any external stimulus, it boosts the photon splitting probability by many orders of magnitude to a value approaching unity.

The central part of our setup is a long on-chip ``telegraph" transmission line terminated by a weak Josephson junction (Fig.~1a, upper panel). Itself made of a chain of 20,000 stronger junctions, the line implements a one-dimensional vacuum
with its wave impedance $Z$ comparable to resistance quantum for Cooper pairs $R_Q =h/(2e)^2 \approx 6.5~\textrm{k}\Omega$, which translates into an effective fine structure constant $\alpha = Z/R_Q$ of order unity~\cite{Kuzmin2019SIT, Martinez2019}. In such a vacuum, microwave photons propagate as sound-like transverse electro-magnetic excitations of the superconducting phase field $\varphi(x,t)$,
described by a quadratic Luttinger liquid-like Lagrangian
\begin{equation}
   L_0=\frac{\hbar v}{4\pi\alpha}\int_{0}^l dx\left[\frac{1}{v^2}\varphi_t^2-\varphi_x^2
    +\frac{1}{\omega_p^2}\varphi_{t,x}^2\right],
\end{equation}
where $v$ is the speed of light in the low-frequency limit and the photon dispersion $\omega(k) = vk/\sqrt{1+(vk/\omega_p)^2}$ has a natural ultra-violet cutoff at $\omega_p/2\pi \approx 20~\textrm{GHz}$, given by the plasma resonance of the chain junctions.
The weak ``impurity" junction presents a non-linear boundary at $x=0$ to the otherwise free field $\varphi (x>0,t)$, which results in the following total system Lagrangian:
\begin{equation}
  L = L_0 +E_J(\Phi) \cos{\varphi(x=0,t)} + \frac{\hbar^2 \varphi_t(x=0,t)^2}{16E_C}.
\end{equation}
The Josephson energy $E_J$ of the impurity junction is tuned by an external flux $\Phi$ using superconducting quantum interference device (SQUID) configuration, and the charging energy $E_C = e^2/2C$ is due to the oxide capacitance $C$. We focus on devices with $E_C \lesssim E_J$, such that the junction mimics a transmon qubit~\cite{koch2007charge} with the resonance at $\omega_0 \approx \left((8E_J E_C)^{1/2} - E_C\right)/\hbar$ and classical damping rate $\Gamma = 4E_C/h\pi\alpha$~\cite{kuzmin2019superstrong}.

\begin{figure*}[t]
	\centering
	\includegraphics[width=0.9\linewidth]{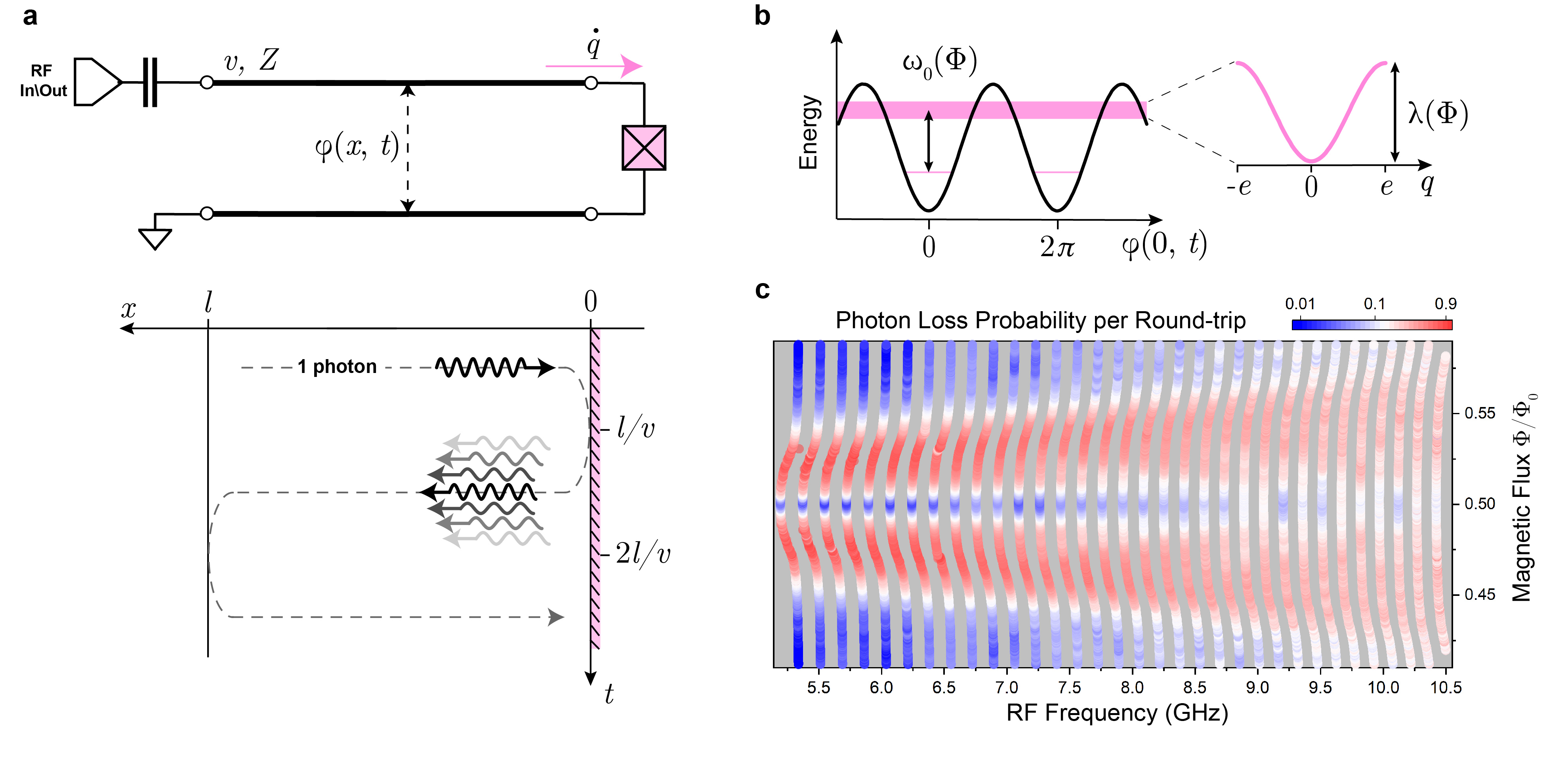}
	
	\caption{(a) Schematic of a telegraph transmission line terminated by a Josephson junction at the right end and weakly coupled to a measurement port at the left end. Device photographs and microwave setup are shown in Ref.~\cite{Kuzmin2019SIT, kuzmin2019superstrong}. The quantum field $\varphi(x,t)$ represents the superconducting phase-difference between the two wires of the line. Lower panel illustrates an inelastic scattering process that splits an incident single photon at $x=0$ in one resonant photon and even number of low-frequency photons. 
	(b) Quantum tunneling of the boundary variable $\varphi(x=0)$ in the periodic Josephson potential renders the energy of the first excited level sensitive to quantum fluctuations of the dynamical charge $q$ at the $x=0$ end of the line. The sensitivity of the ground state can be neglected. (c) The measured positions of standing wave resonances as a function of flux through the split-junction loop in device 3a~\cite{supmat}. The color shows probability to lose a single photon in one round-trip time.
	}

	\label{fig:Fig1}
\end{figure*}

In a harmonic approximation, an incident photon at a frequency $\omega$ would merely scatter off the junction elastically with a phase shift $\delta(\omega) = \arctan((\omega - \omega_0)/\pi\Gamma)$. Inelastic scattering probability due to the conventional self-Kerr non-linearity $\propto \varphi(x=0)^4$ falls into the range $10^{-4} - 10^{-6}$ \cite{kuzmin2019superstrong,supmat} and is hardly measurable. However, a much more efficient non-linearity emerges for $\alpha \gtrsim 1$ from the quantum phase-slip fluctuations across the junction~\cite{matveev2002persistent,rastelli2013quantum}. Namely, tunneling of the phase $\varphi(x=0)$ between the equivalent minima of the Josephson energy renders the energy levels sensitive to the dynamical charge $q$~\cite{averin1985bloch, corlevi2006phase, pechenezhskiy} displaced at the junction end of the transmission line (Fig.~1b). The quantum fluctuations of $q$, unlike those of the boundary phase $\varphi(x=0)$ around a \emph{single}  minimum, are not suppressed at any, even low, frequency. That leads to a profound difference between the conversion processes induced by the self-Kerr versus the phase-slips non-linearity.
The latter opens an infinite number of inelastic scattering channels in the limit of the transmission line's length $l\rightarrow \infty$:  a single incoming photon produces one outgoing photon of a comparable frequency, accompanied by any even number of low-frequency photons. Provided that the phase-slip amplitude is reasonably large, a single incident photon can split with a probability near unity~\cite{supmat}.

The production of low-frequency photons in large quantities has a deep connection to quantum impurity physics~\cite{gull2011continuous}. In fact, for $E_C, \omega_p \rightarrow \infty$, Eqs.~(1),(2) define the boundary sine-Gordon (BSG) quantum impurity model with a critical point at $\alpha = 1$~\cite{gogolin2004bosonization}. The BSG model is important for its integrability property and for describing diverse condensed matter phenomena, from dissipative localization in a periodic potential~\cite{schmid1983diffusion, bulgadaev1984phase} to electron tunneling in Luttinger liquids~\cite{kane1992transport}. The critical dynamics of the field $\varphi$ manifest precisely by inelastic scattering of its bulk excitations -- photons in our case -- off the non-linear boundary~\cite{fendley1994exact}. If the scattering was limited to a mere phase-shift, the boundary could be replaced by a linear one, which would have eliminated interaction effects. Notably, calculating the reflection amplitude $r(\omega)$ as a function of frequency $\omega$ is a difficult task, and it becomes even more so in the presence of the $E_C$-term, which prevents using the exact BSG results. Therefore, measuring $r(\omega)$ would accomplish a useful quantum simulation, which further motivates our experiment. \begin{figure*}
	\centering
	\includegraphics[width=1\linewidth]{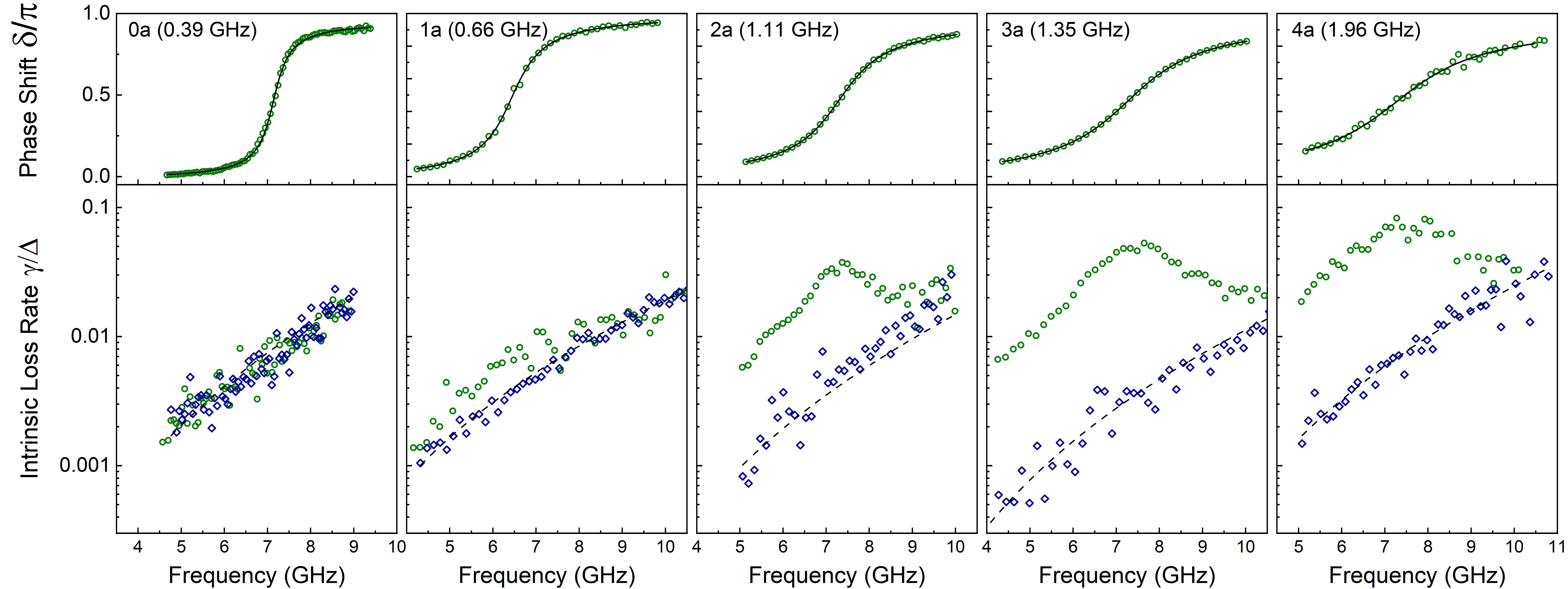}

	\caption{The elastic (top) and inelastic parts (bottom) of the reflection amplitude $r(\omega)$ for the devices with progressively larger charging energy $E_C$ indicated on the plots. In each device, the flux $\Phi$ is tuned such that $\omega_0/2\pi \approx 6.5-7.5~\textrm{GHz}$. The blue markers show data at $\Phi =0$, where the impurity is effectively switched off. The dashed line represents the background dielectric loss inside the transmission line. Device parameters are given in the Table S1 of supplementary material.} 
	
	\label{fig:Fig2}
\end{figure*}

To measure $r(\omega)$ at $x=0$ we introduce a second reflective boundary at $x=l= 6~\textrm{mm}$ in the form of a weakly coupled input/output port. A single photon impinging at the impurity boundary can either scatter elastically with a phase-shift $\delta(\omega)$ or it can split into several left-moving photons (Fig.~1a, lower panel). In both cases, the left-moving photons bounce back at $x=l$ and the process repeats. If the elastic scattering dominates, the two boundaries define a Fabry-P\'erot resonator with a free spectral range $\Delta=v/(2l) \approx 150~\textrm{MHz}$, and the positions of standing-wave mode resonances are linked to 
$\delta (\omega)$.
A rare inelastic event effectively annihilates the photon from a given standing-wave mode as if there is an intrinsic absorption mechanism. Consequently, Fabry-P\'erot resonances would broaden by an amount $\gamma(\omega) \ll \Delta$. The quantities $\delta$ and $\gamma$ are linked to $r$ as $\ln r=2i\delta -2\pi\gamma/\Delta$. Thus, we reduced the scattering experiment in a practically impossible semi-infinite geometry to spectroscopy of Fabry-P\'erot cavity resonances in a finite-size system. As long as the many-body level spacing of the final states is smaller than the scattering rate, our finite-size system behaves similarly to the semi-infinite one. We verified the above condition in our setup (see Fig.4).

Following the previously established rf-spectroscopy technique~\cite{Kuzmin2019SIT}, we identified the frequency and intrinsic linewidth of all standing-wave modes in the $5-10~\textrm{GHz}$ range as a function of flux $\Phi$ (Fig.~1c). The data is taken while populating the modes with much less than one quanta on average, and we checked that the spectroscopic line-shapes remained power-independent. The impurity's resonance has no effect at an integer flux bias $\Phi = 0$, $\Phi_0$ ($\Phi_0 = h/2e$), because then $\omega_0$ is detuned far away towards the plasma cut-off $\omega_p$.  We used data at $\Phi_0 = 0$ to extract the dispersion relation and the value of $Z$, also using the methods from Ref.~\cite{Kuzmin2019SIT}.  
As $\omega_0$ is tuned through the spectrum, multiple modes simultaneously shift by an amount comparable to $\Delta$, signaling the achievement of superstrong coupling condition, $\Gamma \gg \Delta$ \cite{meiser2006superstrong,kuzmin2019superstrong}, required for multi-mode interaction effects. The new effect, though, is an over two orders of magnitude variation of the modes linewidth $\gamma$ with flux. At $\Phi/\Phi_0 \approx 0.475$, the single impurity simultaneously damps over 30 modes, spanning a considerable fraction of the entire energy window. Moreover, the value of $\gamma$ near $5.5~\textrm{GHz}$ is such that photons largely disappear after a single collision with the impurity (Fig.~1c, deep red).

 Mode by mode, we accurately extracted the elastic scattering phase $\delta$ and the intrinsic loss rate $\gamma$ in ten devices with varying parameters (Table S1). The phase $\delta(\omega)$ expectedly winds by $\pi$ across the impurity resonance (Fig.~2, top panels). A fit to the standard oscillator expression provides an accurate estimation of $\Gamma$ and, therefore, $E_C$ (Fig.~2, upper panel). We checked that $\Gamma$ remains flux-independent while growing from $0.6~\textrm{GHz}$ in device 0a to $3.1~\textrm{GHz}$ in device 4a as the impurity junction is fabricated with progressively smaller area (larger $E_C$) \cite{supmat}. The loss rate is flux-independent in device 0a with $E_C = 0.39~\textrm{GHz}$, and it can be explained by the background dielectric absorption in Josephson transmission lines. However, already for $E_C = 0.66~\textrm{GHz}$ in device 1a, there is a noticeable deviation of $\gamma(\omega)$ from the background at $\Phi = 0$, and this deviation rapidly grows with $E_C$ (Fig.~2, lower panels). The anomalous dissipation is maximal for modes located in the $\Gamma$-vicinity of the impurity resonance at $\omega_0$, defined in Fig.~2 as $\delta(\omega_0) =\pi/2$.

Subtracting the background loss of each device from $\gamma(\omega)$, we interpret the remaining rate $\gamma_{\textrm{in}}(\omega)$ as the rate of photon decay due to inelastic scattering at the impurity (Fig.~3). Several properties of $\gamma_{\textrm{in}}$ support our interpretation. The maximal decay rate $\gamma_{\textrm{in}} (\omega = \omega_0)$ grows by an order of magnitude on reducing $\omega_0/2\pi$ by only a few $\textrm{GHz}$. Such a strong frequency dependence of $\gamma_{\textrm{in}}(\omega_0)$ eliminates the possibility of mundane absorption due to either a lossy dielectric or quasiparticle tunneling in the impurity junction. In fact, the growth of $\gamma_{\textrm{in}}(\omega_0)$ at lower frequencies is atypical to materials loss. Furthermore, the rate $\gamma_{\textrm{in}}(\omega_0)$ vanishes in device 1a which features the fastest variation of modes frequency with $\omega_0$ (the sharpest function $\delta(\omega)$ near $\omega = \omega_0$ in Fig.~2). Such an observation eliminates the inhomogeneous broadening mechanism due to slow fluctuations of $\omega_0$ in time. We have also checked that the measured port-coupling is insensitive to flux-bias, and $\gamma_{\textrm{in}}$ is insensitive to increasing the port coupling \cite{supmat}.

Theory supports our interpretation of the anomalous dissipation in terms of photon decay \cite{supmat}. Specifically, for $\alpha > 1$, $\Gamma \ll \omega_0/2\pi$, and $E_C \ll E_J$, the observed photon decay can be quantitatively understood using the following effective phase-slip Hamiltonian,
\begin{equation}
H = \sum_k \hbar\omega_k a_k^{\dagger} a_k + \nu \cos \pi q/e,
\end{equation}
acting at the subset of many-body states with energy near $\hbar\omega_0$. The operators $a_k$ ($a_k^{\dagger}$) annihilate (create) photons at flux-dependent frequencies $\omega_k$, given by positions of the spectroscopic resonances (Fig.~1c) and the effective phase-slip amplitude $\nu$ is proportional to the first Bloch band half-width $\lambda$ of the isolated junction. The dynamical charge $q$ is decomposed over the normal modes according to $q = \sum_k f_k (a_k + a_k^{\dagger})$, where the factors $f_k^2 = (4\pi\Delta/\alpha\omega_k)\times\omega_0^4/\left((\omega_0^2 - \omega_k^2)^2 + (2\pi\Gamma \omega_k)^2\right)$ weight the contribution of individual $k$-modes. In contrast with the Kerr non-linearity, the cosine term in Eq.~3 creates a photon-photon interaction between all the $k$-modes at all even orders. Because $f_k$ is maximal both at $\omega_k = \omega_0$ and at $k=1$, the dominant decay products consist of one near-resonant photon and an even number of low-frequency photons satisfying energy conservation condition. Restricting the calculation to such processes, the inelastic rate for a resonant photon can be found from the Fermi's golden rule:

\begin{equation}
\gamma_{in}(\omega = \omega_0)/\Delta =   \left(\lambda/\omega_0\right)^2 \frac{ \left(\pi\Gamma/\omega_0\right)^{2/\alpha-2}}{2(2/\alpha-1)!\sin (\pi/\alpha) }.
\end{equation}

Within the experimental uncertainty on model parameters, the Eq.~(4) matches the data from all four devices with $\alpha >1$ without adjustable parameters (Fig.~3a, colored bands). Either increasing $E_C$ or reducing $\omega_0$ with the flux-knob exponentially increases $\lambda$, which in turn causes a rapid growth of $\gamma_{\textrm{in}}(\omega_0)$.
The effect of $\alpha$ is weaker, but more complex. In particular, Eq.~4 breaks down for $\alpha \rightarrow 1$, in which case photons are likely produced in the entire frequency range. Devices with $\alpha <1$ exhibit similar, by order of magnitude, decay rates $\gamma_{\textrm{in}}(\omega_0)$, compared to those by devices with $\alpha > 1$ with similar values of $E_C$ (Fig.~3a vs. Fig.~3b). However, a quantitative comparison in case $\alpha <1$ requires more advanced theoretical models than those presently available.

\begin{figure}[htbp]
	\centering
	\includegraphics[width=0.94\linewidth]{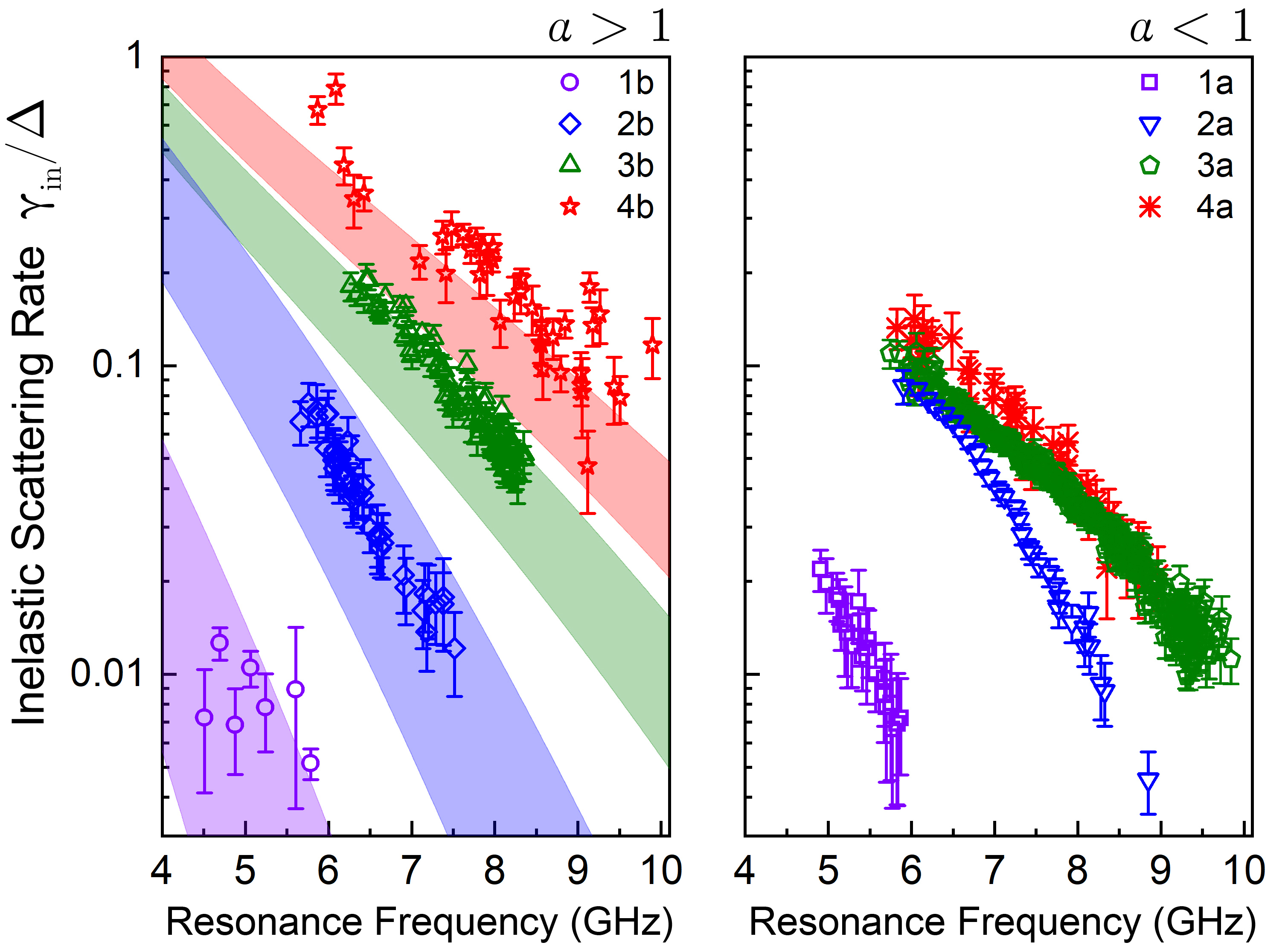}
	\caption{Inelastic scattering rate $\gamma_{\textrm{in}}(\omega_0)/\Delta$ (colored markers) for devices with $\alpha > 1$ (left panel) and $\alpha < 1$ (right panel). The width of theory lines (colored bands) comes from uncertainty in the device parameters. The error bars are the standard errors of $\gamma_{\textrm{in}}/\Delta$ at the resonance. The color code represents nominally identical values of $E_C$.}
	
	\label{fig:Fig3}
\end{figure}

Let us illustrate the large number of decay channels available for a single photon, using an example of mode 47 in device 3a. The flux $\Phi$ is tuned such that $\omega_0/2\pi \approx \omega_{47}/2\pi \approx 6.476~\textrm{GHz}$, and the measured mode linewidth $\gamma_{47} = 11~\textrm{MHz}$.
Using extended spectroscopy data (Fig.~4, left panel), we identified those three-photon and five-photon combinations, whose energy matches $\hbar\omega_{47}/2\pi$ within the half-linewidth $h\times 6.5~\textrm{MHz}$.
This construction reveals a large number of states with a relatively uniform three-photon ($\Delta^{(3)}\approx 1~\textrm{MHz}$, Fig.~4, blue states) and five-photon ($\Delta^{(5)} \approx 50~\textrm{kHz}$, Fig.~4, green states) level spacing. Final states involving higher number of photons are also available and they would form even denser spectrum. We checked that most three-photon states with energies $\hbar(\omega_i + \omega_j + \omega_k)$ couple relatively uniformly, as estimated by their composite weights $f_i f_j f_k$, and the same applies to five-photon states. 
The energy uniformity property comes from a small amount of disorder and dispersion in the single-particle spectrum, which breaks the otherwise massive degeneracy of multi-photon states. These observations justify the treatment of our finite-size transmission line as an infinite one in the derivation of Eq.~4. On reducing the system size (increasing $\Delta$), the many-body spectrum will rapidly become sparse enough to completely suppress the decay. Understanding such energy localization transition in a nearly closed quantum system, originally introduced in the context of Fermi-quasiparticles in a quantum dot~\cite{altshuler1997quasiparticle}, would be a timely extension of our experiment.

\begin{figure}[t]
	\centering
	\includegraphics[width=1\linewidth]{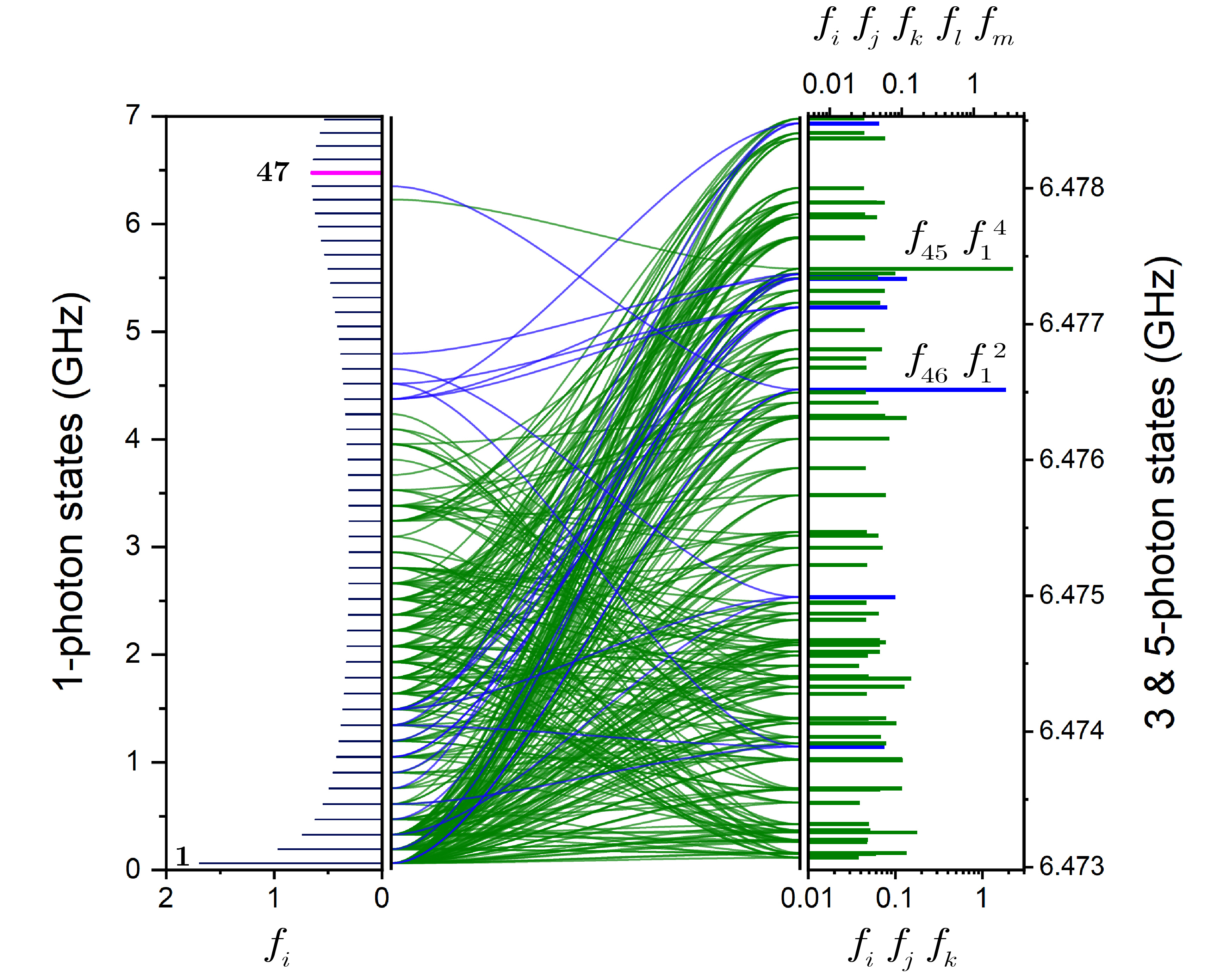}
	\caption{%An example of the states available for the decay of the mode $k=47$ in device 2a for $\omega_0 \approx \omega_{47}$. 
	An example of the many-body states satisfying energy conservation condition for the decay of the mode $k=47$ in device 2a ($\omega_0 \approx \omega_{47}$). The many-body spectrum (right) is obtained by summing all possible combinations of three (blue) and five (green) one-photon frequencies, measured experimentally (left). Each bar's height indicate the one-photon amplitudes $f_k$ (left, see text) and the relative amplitudes of $f_if_jf_k$ and $f_if_jf_kf_lf_m$ of 3-photon and 5-photon states, respectively. The frequency range in the right panel equals to the measured half-linewidth of the $k=47$ mode. The visualization in the central panel illustrates the composition of the multi-photon states from the measured one-photon spectrum. Note the higher weight of the decay channels $\omega_{47} \rightarrow \omega_{46} + 2\omega_{1}$ and $\omega_{47} \rightarrow \omega_{45} + 4\omega_{1}$ involving the lowest frequency mode at $\omega_1/2\pi = 63~\textrm{MHz}$.}
	
	\label{fig:Fig4}
\end{figure}

In summary, a quantum phase-slip center in high-impedance superconducting waveguides can split a single incident photon into a large number of lower-energy photons with probability near unity. Inserting such an efficient inelastic scattering center inside a closed Fabry-P\'erot resonator makes the photon lifetime comparable to the round-trip time, in which case the standing-wave resonances are damped by the photon-photon interaction to the degree prohibiting the use of free-photon description of the quantum electromagnetic field in the resonator.
Notably, the underlying regime of extreme non-linearity in circuit quantum electrodynamics opens the door to simulating strongly-correlated phenomena, including superconductor-insulator transitions in one-dimensional systems~\cite{bard2018decay, wu2018theory, houzet2019microwave}.

Looking ahead, our circuit spectroscopy technique can be applied to simulate important quantum impurity models. For instance, reducing the junction size (increasing $E_C$) would implement the BSG-model. Shunting the weak junction by an inductance would implement a spin-boson model, related to Anderson and Kondo models~\cite{garcia2008quantum, le2012kondo}, in which case a large inelastic scattering cross-section was predicted near the Toulouse point~\cite{goldstein2013inelastic}. Furthermore, rapidly switching the impurity on and off with the flux knob would induce controlled out-of-equilibrium dynamics. The present measurement of $r(\omega)$ already implements an example analog quantum simulation of a many-body quantity which is non-trivial to calculate otherwise.
We verified the simulation outcome in the parameter regime available to analytical calculations (Fig.~3, left panel). The rest of data (Fig.~3, right panel) represents a unique quantum resource for benchmarking numerical methods~\cite{bravyi2017complexity}.\\ %and possibly testing noisy digital quantum computers, whose focus will likely include quantum impurity problems~\cite{bravyi2017complexity}.\\
\\
We acknowledge useful discussion with I. Protopopov and J. Sau and funding by the US ARO MURI program, US-Israel BSF, and US DOE contract DE-FG02-08ER46482 (LG).

%\bibliography{SuperconductingCircuits}

\providecommand{\noopsort}[1]{}\providecommand{\singleletter}[1]{#1}%

\end{document}